\documentclass[final,times,pdflatex,twocolumn,showpacs,superscriptaddress,nofootinbib]{elsarticle}
\usepackage{bm,amsmath,amssymb}
\usepackage{graphicx}
\usepackage{subfigure}
\usepackage{color}
\usepackage{soul}
\usepackage{hyperref}
\usepackage{multirow}
\usepackage{lineno}
\usepackage{footnote}
\setulcolor{blue}
\setstcolor{red}
\sethlcolor{yellow}
\newcommand \beq {\begin{equation}}
\newcommand \eeq {\end{equation}}
\newcommand \beqa {\begin{eqnarray}}
\newcommand \eeqa {\end{eqnarray}}
\def\lsim{\raise0.3ex\hbox{$<$\kern-0.75em\raise-1.1ex\hbox{$\sim$}}}

\journal{Physics Letters B}

\begin{document}

\begin{frontmatter}

\title{Charm degrees of freedom in hot matter from lattice QCD}

\author[affiliation1]{A. Bazavov}
\author[affiliation2,affiliation3]{~D. Bollweg}
\author[affiliation4]{~O. Kaczmarek}
\author[affiliation4]{~F. Karsch}
\author[affiliation2]{~Swagato Mukherjee}
\author[affiliation2]{~P. Petreczky}
\author[affiliation4]{~C. Schmidt}
\author[affiliation4]{~~~~~~~~Sipaz Sharma}
\address[affiliation1]{Department of Computational Mathematics, Science and Engineering and Department of Physics and Astronomy, Michigan State University, East Lansing, MI 48824, USA}
\address[affiliation2]{Physics Department, Brookhaven National Laboratory, Upton, New York 11973, USA}
\address[affiliation3]{RIKEN-BNL Research Center, Brookhaven National Laboratory, Upton, New York 11973, USA}
\address[affiliation4]{Fakult\"at f\"ur Physik, Universit\"at Bielefeld, D-33615 Bielefeld, Germany}

%------------------------------------------------------------------------------------
%       abstract
%------------------------------------------------------------------------------------

\begin{abstract}

We study the nature of charm degrees of freedom in hot strong interaction matter
by performing lattice QCD calculations of 
the second and fourth-order cumulants of charm  fluctuations, and their correlations with net baryon number, electric charge and strangeness  fluctuations. We show that below the chiral crossover temperature 
thermodynamics of charm can be very well understood in terms of charmed hadrons. 
Above the chiral transition charm quarks show up as new degrees of
freedom contributing to the partial charm
pressure. However, up to temperatures as high as 175 MeV charmed hadron-like excitations
provide a significant contribution to the partial charm pressure.

\end{abstract}


\end{frontmatter}

\section{Introduction}
\label{introduction}

It is now well established that  strong interaction matter at vanishing baryon chemical potential undergoes restoration of the spontaneously broken chiral symmetry via a crossover transition since the small yet non-vanishing up and down quark masses result also in the explicit breaking of the ${SU(2)_L\times SU(2)_R}$ chiral symmetry group. This chiral crossover transition occurs at a pseudo-critical temperature, %${T_{pc}}$, which calculated using lattice QCD formalism is 
${T_{pc}=156.5\pm1.5}$ MeV \cite{HotQCD:2018pds}. However, the deconfinement aspect of the transition is 
not well understood in QCD with light dynamical quarks. Ultimately, the deconfinement can
be related to the nature of the underlying degrees of freedom in the hot matter. We expect that with increasing
temperatures hadronic excitations become broader and may have masses different from the vacuum ones.
This is the case for the pseudo-scalar mesons \cite{Bala:2023iqu} and the vector mesons, e.g. the $\rho$-meson \cite{Rapp:1997fs,Rapp:1999ej,Hohler:2013eba,Jung:2016yxl}. 
There is some evidence that also charmed hadrons get modified with increasing temperature \cite{Kelly:2018hsi,Aarts:2022krz,Aarts:2023nax}. At sufficiently high temperatures the hadronic excitations will become too broad and not well defined,
and partonic excitations will be the dominant ones.

Fluctuations of conserved charges and correlations among different conserved
charges can be used to understand the relevant degrees of freedom in the hot
matter. Precision lattice QCD calculations showed that the Hadron Resonance Gas (HRG) \cite{Fiore:1984yu}
provides a good description of the fluctuations and correlations of conserved
charges in the light and strange quark
sector below the chiral crossover \cite{Bellwied:2015lba,Bollweg:2021vqf}. 
There are indications that the same is true for charm fluctuations
and charm baryon number correlations
\cite{BAZAVOV2014210}. However, because of the large statistical
errors no firm conclusion 
could be drawn in the earlier
studies about the onset for the presence of
charm quark degrees of freedom
\cite{BAZAVOV2014210}.
At high temperature the fluctuations and correlations of conserved charges 
can be well understood in terms of a quark gas \cite{Bellwied:2015lba,BAZAVOV2014210,Bazavov:2013uja,Ding:2015fca}.

For understanding of the  spectrum 
and elliptic flow of charmed hadrons it is important to know how these
hadrons are formed and whether charmed hadron states can exist also above the chiral crossover temperature
 \cite{Ravagli:2007xx,Mannarelli:2005pz}. Using previously obtained lattice
 QCD data on charm baryon number correlations \cite{BAZAVOV2014210} 
 it has been argued that 
 charmed hadron-like excitations can exist above $T_{pc}$
\cite{Mukherjee:2015mxc}.

The aim of this letter is to clarify the nature of charm degrees
of freedom in the vicinity of $T_{pc}$ using high precision calculations of 
charm fluctuations and correlations.
We test to what extent the HRG model
can describe the thermodynamics of charm below the chiral crossover.
Furthermore, we probe the onset of deconfinement and
establish the existence of charmed hadronic excitations above $T_{pc}$.

\section{Details of the lattice QCD calculations}
\label{sec:lat}
In order to study the nature of the charm degrees of freedom we calculate
so-called generalized susceptibilities, i.e. derivatives of the QCD pressure ($P$)
with respect to chemical potentials of net baryon number ($B$), electric charge ($Q$), strangeness ($S$)
and charm ($C$),
\begin{equation} 
{\chi^{BQSC}_{klmn}=\dfrac{\partial^{(k+l+m+n)}\;\;[P\;(\hat{\mu}_B,\hat{\mu}_Q,\hat{\mu}_S,\hat{\mu}_C)\;/T^4]}{\partial\hat{\mu}^{k}_B\;\;\partial\hat{\mu}^{l}_Q\;\;\partial\hat{\mu}^{m}_S\;\;\partial\hat{\mu}^{n}_C}}\bigg|_{\vec{\mu}=0} \; ,
\label{eq:chi}
\end{equation}
in lattice QCD. We introduced a dimensionless notation for chemical potentials, 
${\hat{\mu}_X = \mu_X/T}$, with $X \in \{B, Q, S, C\}$.

We performed lattice QCD calculations in (2+1)-flavor QCD using the Highly Improved Staggered Quark (HISQ) action \cite{Follana:2006rc} for physical strange
quark mass, $m_s$, and light quark mass, $m_l=m_s/27$. The latter corresponds
to a pion mass of $140$ MeV in the continuum limit. We consider the temperature range from 145 MeV to 175 MeV.
To fix the lattice spacing the 
$f_K$ scale setting from Ref. \cite{Bollweg:2021vqf}
is used. The values of lattice strange quark mass are taken from
Ref. \cite{HotQCD:2014kol}. 
We use a set of gauge field configurations generated on lattices of size $32^3\times 8$ and used in 
earlier studies of the HotQCD collaboration \cite{Bollweg:2021vqf}. The charm quarks have been treated in quenched
approximation, which can be justified since the charm quark mass, $m_c$, is quite large,
and earlier lattice calculations showed that the influence of 
dynamical charm quarks can be neglected
in the temperature range of interest for our current analysis \cite{Borsanyi:2016ksw}. The HISQ action is very well suited for the study of charm
quarks \cite{Follana:2006rc}. Discretization effects related  to  the  charm  quarks  can  be reduced  by  using a mass-dependent coefficient in the HISQ action which eliminates 
${\cal O}((a m_c)^4)$ tree level lattice artifacts \cite{Follana:2006rc,MILC:2010pul}.
\begin{table}[t]
\centering
\begin{tabular}{|c|c|c|c|}
\hline
~&&\multicolumn{2}{c|}{$am_c$}  \\
\hline
$\beta$ & $T$[MeV] & LCP[a] & LCP[b] \\
\hline
6.315 & 145.1 & 1.04112 &0.892231\\
6.354 & 151.1 & 0.97025 &0.857304\\
6.390 & 156.9 & 0.91534 &0.816144\\
6.423 & 162.4 & 0.87069 &0.787450\\
6.445 & 166.1 & 0.84320 &0.765223\\
6.474 & 171.2 & 0.80920 &0.742996\\
6.500 & 175.8 & 0.78059 &0.723946\\
\hline       
\end{tabular} 
\caption{\label{tab:amc} 
The lattice gauge coupling $\beta=10/g_0^2$, the corresponding temperature values and the bare 
charm quark masses for LCP[a] and LCP[b]. 
}
\end{table}
We calculated all generalized susceptibilities
involving charm up to fourth order.
The calculation of $\chi^{BQSC}_{klmn}$ involves derivatives of the pressure and on the lattice this is achieved by the unbiased stochastic estimation of various traces -- consisting of inversions and derivatives  of the fermion matrices ($D$) -- using the random noise method \cite{Mitra:2022vtf}. In particular, $500$ random vectors have been used to calculate various traces per configuration, except for Tr ${\big(D^{-1}\frac{\partial D}{\partial \mu}\big)}$ -- which turned out to be particularly noisy. Therefore, $2000$ random vectors have gone into its calculation. 
We used two different lines of constant physics (LCPs) to tune the charm-quark mass. The first LCP corresponds to keeping the spin-averaged charmonium mass, ${(3m_{J/\psi}+m_{\eta_{c\bar{c}}})/4}$ fixed to its physical value. We calculated
the $J/\psi$ and $\eta_c$ masses using the zero temperature lattices generated
for the study of chiral crossover temperature and equation of state \cite{HotQCD:2014kol}. 
As in Ref. \cite{BAZAVOV2014210}, we fitted the corresponding values of the bare charm quark mass $am_c$ with renormalization group inspired
form to obtain $am_c$ as function of the inverse gauge coupling, $\beta=10/g_0^2$. In Table \ref{tab:amc} we
give the values of the charm quark masses obtained through this procedure used in our calculations
for different $\beta$ as well as the temperature values.
The second LCP  is defined by the physical charm to strange quark mass ratio, $m_c/m_s=11.76$ \cite{ParticleDataGroup:2022pth}. Results based on the above two LCPs will henceforth contain subscripts [a] and [b], respectively. Results without any of these subscripts will correspond to LCP[b]. For $\beta>6.75$, the ratio $m_c/m_s$ on LCP[a] converges to the
experimental value used for LCP[b] \cite{Sharma:2024ucs}. 
At $\beta$-values close to the pseudo-critical $\beta$ of $N_\tau=8$ lattices used here, this ratio varies, however, by (10-15)\% as can be 
seen in Tab. \ref{tab:amc}. As a consequence, the charmed hadron masses calculated
on different LCPs differ. For instance, at $\beta=6.39$, corresponding
to $T=156.9$~MeV, we find $m_{\eta_{c\bar{c}}}=3.077(7)$~GeV on LCP[a] and 
2.879(2)~GeV on LCP[b], {\it i.e.} the larger than physical quark mass ratio
on LCP[a] results in a larger charmonium mass on that LCP. The same holds for the open charm hadron masses.

As stated above our aim is the high statistics lattice QCD calculations
of the generalized susceptibilities involving charm.
The generalized susceptibilities involving charm have been first
studied in Ref. \cite{BAZAVOV2014210} using $32^3\times 8$ lattices
in (2+1)-flavor. We extend this study 
as well as the analysis of charmed
degrees of freedom \cite{Mukherjee:2015mxc}
in two significant ways. 
We included in the analysis two temperatures below the crossover temperature and increased the statistics by a factor (60-70) 
in the vicinity of $T_{pc}$ and 
a factor 20 at $T\simeq 175$~MeV.
The smallest temperature used in 
Ref.~\cite{Bazavov:2014yba} was 
$T=157$ MeV, which is too high to test the range of validity of HRG model 
calculations. 
We have used approximately one-third of the available (2+1)-flavor HISQ configurations generated by the HotQCD collaboration \cite{Bollweg:2021vqf}.
At the highest two temperatures, i.e., at $T=171.6$ MeV
and $T=176.7$ MeV we also performed calculations on $48^3 \times 12$ lattices. At these two temperatures we used 36078 and 39080 gauge configurations,
respectively.

\begin{figure}
	\centering 
\includegraphics[width=0.4\textwidth, angle=0]{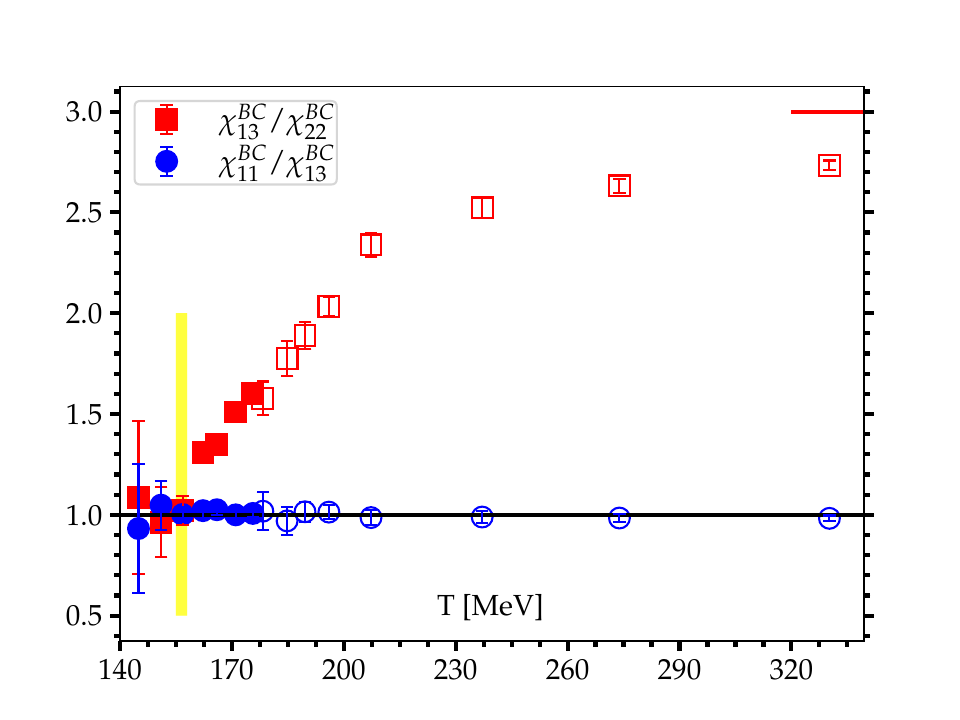}
\caption{The ratios of different baryon-charm fluctuations as functions of temperature. The open symbols
represent the results from Ref. \cite{BAZAVOV2014210}.
The yellow band represents $T_{pc}$ with its uncertainty.
The red solid line is the ideal charm quark gas limit of the ratio $\chi_{13}^{BC}/\chi_{22}^{BC}$.}
	\label{fig:chiBC22}%
\end{figure}

\section{Generalized charm susceptibilities and hadron resonance gas model}

In a non-interacting HRG model the QCD pressure can be written as the sum of
the partial pressures of hadrons
carrying open charm degrees of freedom and the partial pressure of hadrons with no charm.
Furthermore, the partial pressure of charmed hadrons can be written as the sum of partial pressures
of charmed mesons and charmed baryons.
\begin{equation}
{P_C(T,\vec{\mu})=P_M^C(T,\vec{\mu})+P_B^C(T,\vec{\mu})} \text{ .}
\label{eq:P}
\end{equation}
As the masses of charmed mesons and baryons are much larger than the temperature range of interest, one can use Boltzmann statistics 
and write $P_M^C$ and $P_B^C$ in the following form \cite{BAZAVOV2014210}:
\begin{gather}
\begin{aligned}
	P_{B/M}^C(T,\vec{\mu})&=\dfrac{1}{2\pi^2}\sum_{i\in \text{C-B/M}}g_i \bigg(\dfrac{m_i}{T}\bigg)^2K_2(m_i/T)\\
  &\times\text{cosh}(B_i\hat{\mu}_B+Q_i\hat{\mu}_Q+S_i\hat{\mu}_S+C_i\hat{\mu}_C) \text{ .}
	\label{eq:McBc}
\end{aligned}
\end{gather}
Here $B=0$ gives the meson pressure, $P_M^C$,
   and $B= \pm 1, \pm 2, ...$ gives the baryon pressure, $P_B^C$; ${K_2(x)}$ is a modified Bessel function of the second kind; the summation is over all charmed baryons/mesons with masses given by ${m_i}$; $g_i$ denotes the degeneracy factors of states with identical mass and quantum numbers.
   In the HRG phase, generalized susceptibilities introduced in Eq.~\eqref{eq:chi} are calculated by making use of the partial pressure expressions introduced above in Eq.~\eqref{eq:McBc}. In particular, for the calculation of generalized susceptibilities at vanishing chemical potential in the charm sector, it suffices to replace the QCD pressure, $P$, with the partial charm pressure, $P_C$. The final expression for $\chi^{BQSC}_{klmn}$ takes the following form:

\begin{equation}
\chi^{BQSC}_{klmn}=\dfrac{1}{2\pi^2}\sum_{i\in \text{C-H}}g_i \bigg(\dfrac{m_i}{T}\bigg)^2K_2(m_i/T)\; B^kQ^lS^mC^n \text{,}
	\label{eq:chi_HRG}
\end{equation}
where the sum is over all charmed hadrons. 
According to Eq.~\eqref{eq:chi_HRG}, particles with different quantum numbers contribute with different 
weights to different generalized susceptibilities.
Fore example, for $\chi_{13}^{BC}$  the contribution of a particle with $B=2$ will contain a factor $2$, whereas for $\chi_{22}^{BC}$ it will contain a factor $2^2$. On the other hand particles with $B=1$
contribute with the same weight to $\chi_{13}^{BC}$ and $\chi_{22}^{BC}$.
Therefore, below $T_{pc}$ one thus would expect to find $\chi_{13}^{BC}/\chi_{22}^{BC}<1$, if there is significant contribution from dibaryons and $\chi_{13}^{BC}/\chi_{22}^{BC} \simeq 1$
otherwise. 

\begin{figure}
	\centering 
	\includegraphics[width=0.4\textwidth, angle=0]{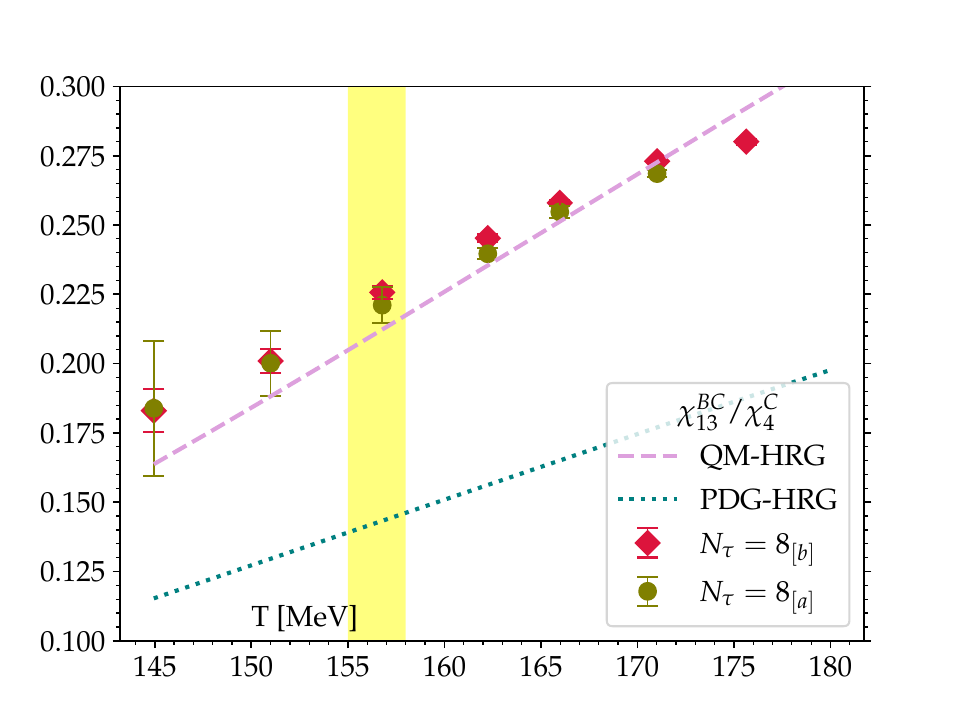}	
	\caption{The ratio $\chi_{13}^{BC}/\chi_4^C$ as a function of temperature obtained for LCP[a] and LCP[b].
	Also shown is the result obtained in PDG-HRG and QM-HRG model calculations.
	The yellow band represents $T_{pc}$ with its uncertainty.
	} 
	\label{fig:chiBC13}%
\end{figure}

For large value of the argument, $K_2(m_i/T)$ is exponentially suppressed. Therefore,
the contribution of multi-charm  baryons is exponentially small, and effectively only the $|C|=1$ sector contributes
to the pressure. 
This means that $\chi_2^C=\chi_n^C=P_C(T,\vec{\mu})$, for $n$ even, and
$\chi_{11}^{BC}=\chi_{1m}^{BC}=P_B^C$, for $m$ odd. 
We also note that these relations should hold also for $T>T_{pc}$,
where charm quarks are expected to be the relevant degrees of freedom, because for temperatures a few times $T_{pc}$, the Boltzmann approximation also works for an ideal massive charm quark gas,  see discussion in the next section. 
We find that these
relations are indeed well satisfied in the temperature range used in our
calculations as shown in Fig. \ref{fig:chiBC22}. 
From Fig. \ref{fig:chiBC22} we also see that 
$\chi_{13}^{BC}/\chi_{22}^{BC}$ close to one up to the crossover temperature.
As discussed above this is expected in a hadron gas if the contribution of
$|B|=2$ sector is small. Our lattice results cannot rule out a small
contribution from charmed dibaryons given the statistical errors.
In Fig.~\ref{fig:chiBC22} we also show the earlier
lattice QCD results as open symbols for $T>176$ MeV \cite{BAZAVOV2014210}. At lower temperatures our results
agree with those of Ref.~\cite{BAZAVOV2014210} within the large errors of the latter. The present results
have much smaller errors.
Just above the chiral
crossover temperature the HRG description breaks down and the ratio approaches
a value which at 330 MeV is only 10\% below the value of an ideal charm quark gas.

In Fig.~\ref{fig:chiBC13} we show the 
ratio $\chi_{13}^{BC}/\chi_4^C$, which is a proxy for the ratio of the charmed baryon pressure and the total
charm pressure. We compare our lattice QCD results with HRG model predictions where we include all charmed hadrons listed by the Particle Data Group \cite{ParticleDataGroup:2022pth} (PDG-HRG). 
As can be seen, the PDG-HRG under-predicts the lattice data significantly. This is not
surprising since it was pointed out
already in Ref. \cite{BAZAVOV2014210} that many charmed baryons predicted in quark model \cite{Ebert:2011kk} as well as lattice QCD calculations 
\cite{HadronSpectrum:2012gic}
are missing in PDG tables. There are also missing charmed meson states in PDG-HRG. However, their number is significantly smaller
\cite{BAZAVOV2014210}.
When including all the missing hadron states, using the spectrum obtained in quark model calculations  
\cite{Chen:2022asf,Ebert:2011kk,Ebert:2009ua}, in the HRG model, 
we obtain very good agreement between the lattice QCD results and the quark model extended hadron resonance gas (QM-HRG) for $\chi_{13}^{BC}/\chi_4^C$ for $T \le 170$ MeV. We do not consider dibaryon contribution to QM-HRG since there is no
clear evidence for such states.
The heaviest state in our QM-HRG data set has a mass of about $4.4~\text{GeV}$. In our previous work \cite{BAZAVOV2014210}, we showed that QM-HRG calculations based on QM-HRG data set containing states with masses less than $3.5~\text{GeV}$ agree with the complete QM-HRG model results to better than 1\%.
Furthermore, as one can also see in Fig. \ref{fig:chiBC13} the ratio $\chi_{13}^{BC}/\chi_4^C$
is not very sensitive to the choice of the LCP for the charm mass. Thus our present findings are in agreement
with the observations of 
Ref.~\cite{BAZAVOV2014210} that additional charmed baryon states are needed to explain
the lattice QCD results on generalized susceptibilities, but now this claim is on more solid numerical footings, because
now we have two more data points below $T_{pc}$, and our
statistical errors are smaller.
The apparent agreement between the lattice QCD results and HRG
also for $T>T_{pc}$, that can be 
seen in
Fig. \ref{fig:chiBC13}, is somewhat accidental. It
is related to the fact that at high temperatures
$\chi_{13}^{BC}/\chi_4^C$ approaches 1/3. In fact, any other ratio with a smaller ideal charm quark gas limit, e.g., $\chi_{22}^{BC}/\chi_4^C$, shows a clear departure from the HRG description at $T_{pc}$ (see also the data in the $QC$ sector \cite{Sharma:2024ucs}). 
Thus, the HRG description description breaks down for $T>T_{pc}$, and we also see that 
for $T_{pc} <T <350$ MeV the ratio  $\chi_{13}^{BC}/\chi_{22}^{BC}$ moves from the 
hadron gas to the quark gas expectations. In the next section we discuss 
the implication of this finding for effective charm degrees of freedom.

\begin{figure}[ht]
\includegraphics[width=0.48\textwidth]{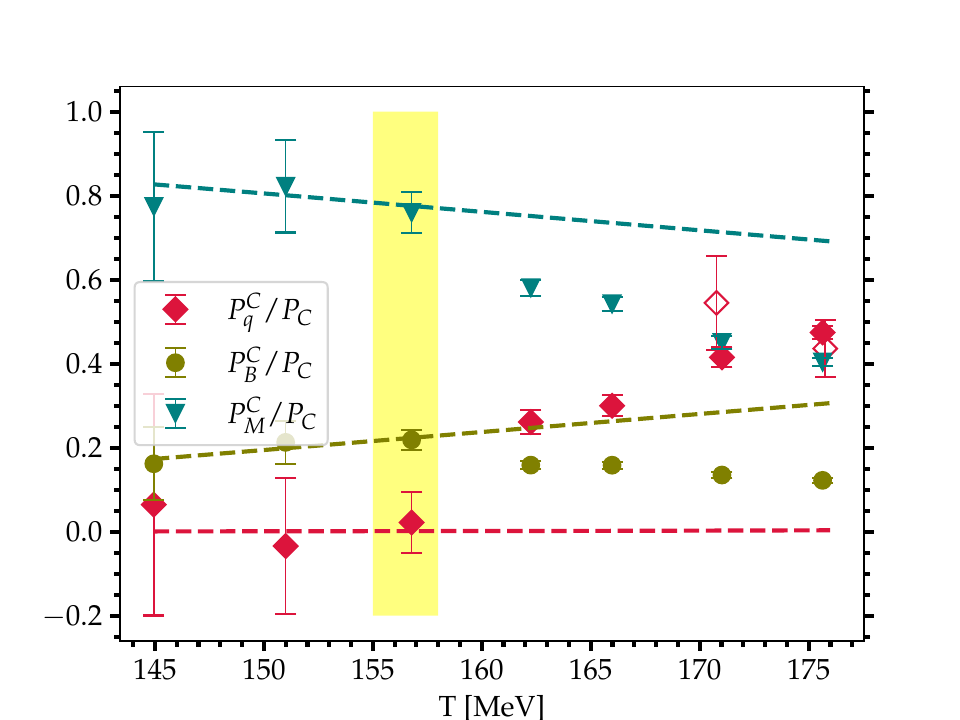}
\caption{Partial pressures of charmed mesons, charmed baryons and charm quarks as functions of temperature. All three observables have been normalized to the total partial charm pressure. The dashed lines show corresponding results obtained from the QM-HRG model. The open symbols show the results for $N_{\tau}=12$ lattices, see text. The yellow band represents $T_{pc}$ with its uncertainty.} 
\label{fig:quasi}
\end{figure}
\begin{figure*}[ht]
\includegraphics[width=0.48\linewidth]{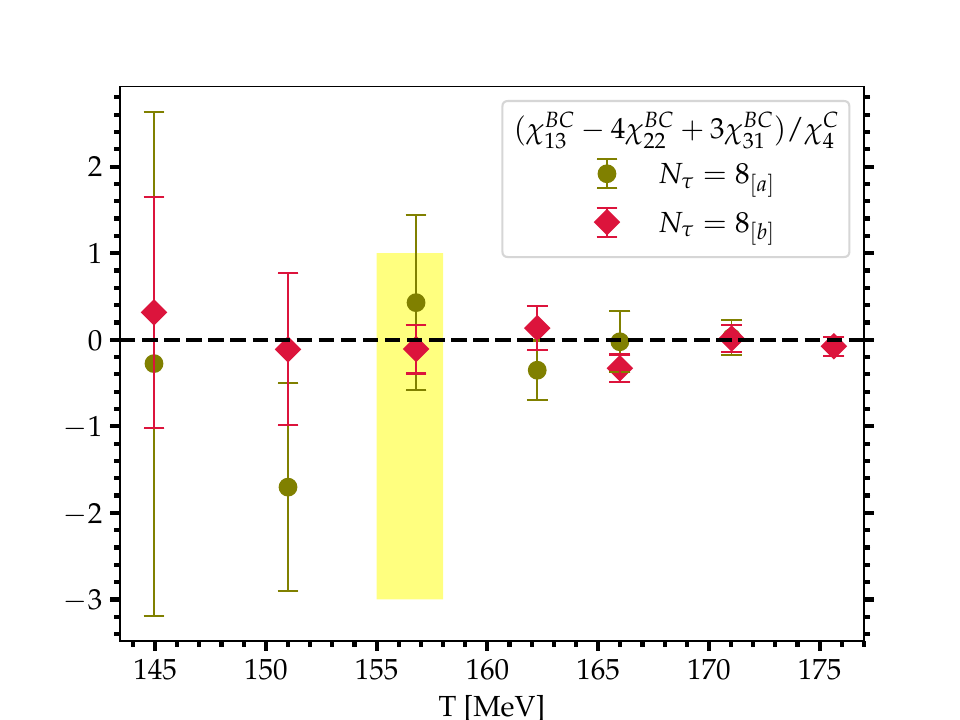}
\includegraphics[width=0.48\linewidth]{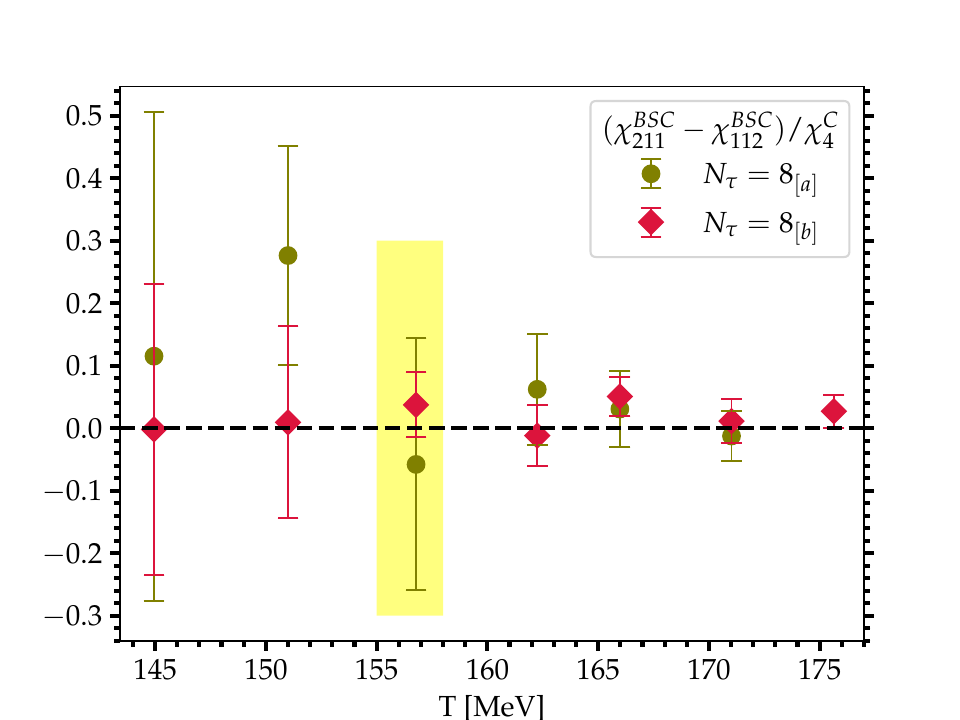}
\caption{Constraints $c_1$ and $c_4$
on the quasi-particle model given by Eq. \ref{eq:Pmodel}, see text
and Eqs.~\ref{c1} and Eq. \ref{c4}, respectively.
The dashed lines indicate that these constraints are explicitly fulfilled in
non-interacting HRG models. 
The yellow band represents $T_{pc}$ with its uncertainty. 
}
\label{fig:const}
\end{figure*}

\section{Charm degrees of freedom below and above $T_{pc}$}

As shown in the previous section the simple hadron gas model, given by Eqs. \ref{eq:P} and \ref{eq:McBc},
breaks down for $T>T_{pc}$. Therefore, following Ref.~\cite{Mukherjee:2015mxc} 
we extend this model by allowing the presence of partial charm quark pressure and by treating
the charmed baryon and charmed meson sectors as  quasi-particle excitations,
\begin{align}
    P_C(T,\vec{\mu})=P_M^C(T,\vec{\mu})+P_B^C(T,\vec{\mu})+\nonumber\\
    P_q^C(T) \cosh\bigg(\dfrac{2}{3}\hat{\mu}_Q+\dfrac{1}{3}\hat{\mu}_B+\hat{\mu}_C\bigg)\; ,
    \label{eq:Pmodel}
\end{align}
where the last term corresponds to the charm quark partial pressure. Here
we explicitly give the dependence 
on chemical potentials, using Boltzmann approximation, and 
$P_q^C(T)$ being a function of temperature only.
At very high temperature we can also write $P_q^C(T)=\frac{3}{\pi^2}\bigg(\frac{m_c}{T}\bigg)^2K_2(m_c/T)$, where
$m_c$ is the mass of a quasi-particle with quantum numbers of charm quark.
Using Eqs. (\ref{eq:McBc}) and (\ref{eq:Pmodel}) we can express the partial pressures of quasi-particles with  quantum numbers of charm quarks, 
charmed baryons
and charmed mesons, respectively. For $\vec{\mu}=0$ we 
express these partial pressures in terms of generalized susceptibilities as follows,
\begin{align}
	P_{q}^{C}&=9(\chi^{BC}_{13}-\chi^{BC}_{22})/2\; , 
	\label{eq:partial-quasiq}\\
	P_{B}^{C}&=(3\chi^{BC}_{22}-\chi^{BC}_{13})/2\; , 
	\label{eq:partial-quasiB}\\
	P_{M}^{C}&=\chi^{C}_{4}+3\chi^{BC}_{22}-4\chi^{BC}_{13} \; .
	\label{eq:partial-quasiM}
\end{align}
Using lattice QCD results on the generalized susceptibilities in the above expressions we estimate
the different partial pressures
and normalize them by dividing with
the total partial charm pressure,
$P_C=\chi^C_4$. Results are 
shown in Fig.~\ref{fig:quasi}. 
Corresponding results obtained in
QM-HRG model calculations are shown
as dashed lines in this figure.
Obviously the quark pressure is zero in this model. As can be seen, the HRG works well up to $T_{pc}$. Above $T_{pc}$, however, the charmed
baryon and charmed meson partial pressures drop below the HRG results.
At the same time the quark pressure
becomes non-zero just above $T_{pc}$. 
These results may be taken as evidence for partial melting of the hadron-like 
states and the liberation of quark 
degrees of freedom at $T_{pc}$.
For the highest two temperatures we also show our results for the charm quark pressure
obtained on $N_{\tau}=12$ lattices, which agree with the $N_{\tau}=8$ results within errors, indicating that our main
conclusion is not affected by lattice artifacts.

As was pointed out in Ref.~\cite{Mukherjee:2015mxc}, 
if the model given by
Eq. (\ref{eq:Pmodel}) takes care of 
all relevant degrees of freedom below
and above $T_{pc}$, 
then there are four constraints that the generalized
susceptibilities up to fourth order must satisfy:
\begin{align}
c_1 \equiv \chi_{13}^{BC} - 4\chi_{22}^{BC} + 3\chi_{31}^{BC} = 0 \label{c1},\\
c_2 \equiv 2\chi_{121}^{BSC} + 4\chi_{112}^{BSC} + \chi_{22}^{SC} - 
2\chi_{13}^{SC} + \chi_{31}^{SC} = 0 , \label{c2} \\
c_3 \equiv 3\chi_{112}^{BSC} + 3\chi_{121}^{BSC} - \chi_{13}^{SC} + 
\chi_{31}^{SC}=0 , \label{c3} \\
c_4 \equiv \chi_{211}^{BSC} - \chi^{BSC}_{112}=0 . \label{c4}
\end{align}
The lattice results in Ref. \cite{BAZAVOV2014210} were consistent with these constraints. 
However, because of the large errors it was not possible to test the model
for $T<200$ MeV. With our new lattice data one can show that these constraints are
clearly satisfied in the vicinity of $T_{pc}$ within errors. As an example in Fig. \ref{fig:const} we show
the constraints $c_1$ (left)  and $c_4$ (right). 
The observables defined by Eq. \ref{c1} and Eq. \ref{c4} are in fact related to the partial pressures
of charm diquarks, $P_d^C$, and charm-strange diquarks, $P_{ds}^C$. Namely we can write $P_d^C=-9 c_1/2$ and $P_{ds}^C=9 c_4/2$.
While the current data on $c_1$ and $c_4$ are consistent with not having a diquark
contribution to the pressure at $T>T_{pc}$, the current bounds coming from the
data on $c_1$ and $c_4$ are still too weak to rule out the existence of diquarks above
$T_{pc}$ completely.

\begin{figure}[t]
\includegraphics[width=0.4\textwidth]{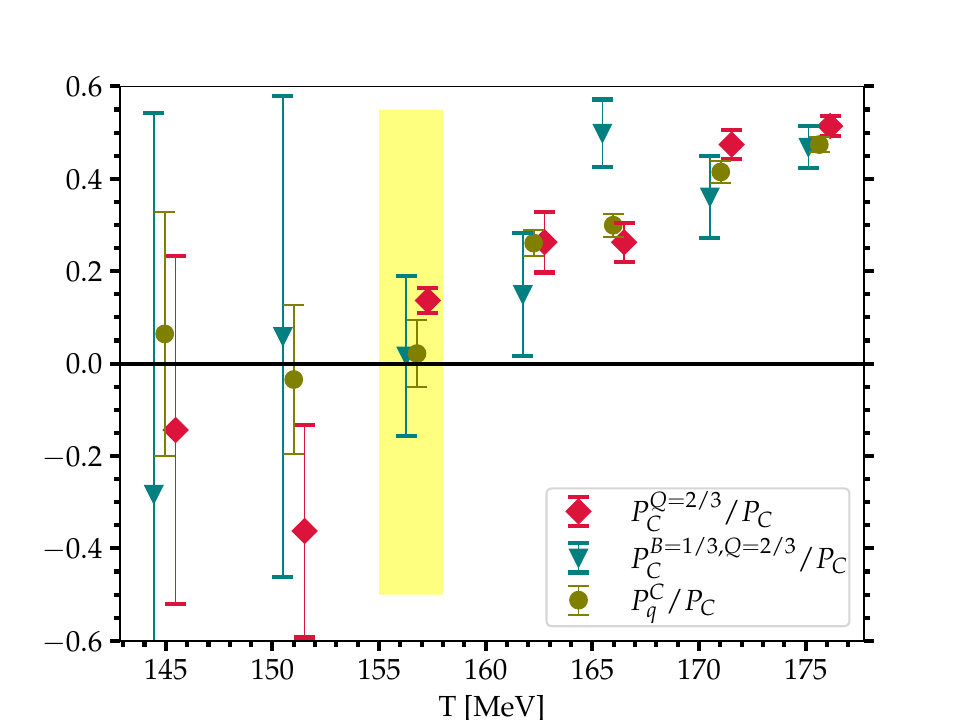}
\caption{The partial pressures 
of quasi-particles carrying
(i) baryon number $1/3$ ($P_q^C$), (ii) electric charge
$Q=2/3$ ($P_C^{Q=2/3}$), and
(iii) ($B=1/3,\ Q=2/3$)  ($P_C^{B=1/3,Q=2/3}$), respectively. All three observables have been normalized to the total partial charm pressure.
The figure confirms that these three observables project onto the
same quasi-particle sector.
The yellow band represents $T_{pc}$ with its uncertainty. Note that the T-coordinates of case (ii) and case (iii) are shifted by $\pm 0.51$ MeV respectively.
}
\label{fig:Pqc}
\end{figure}

\begin{figure}[t]
\includegraphics[width=0.4\textwidth]{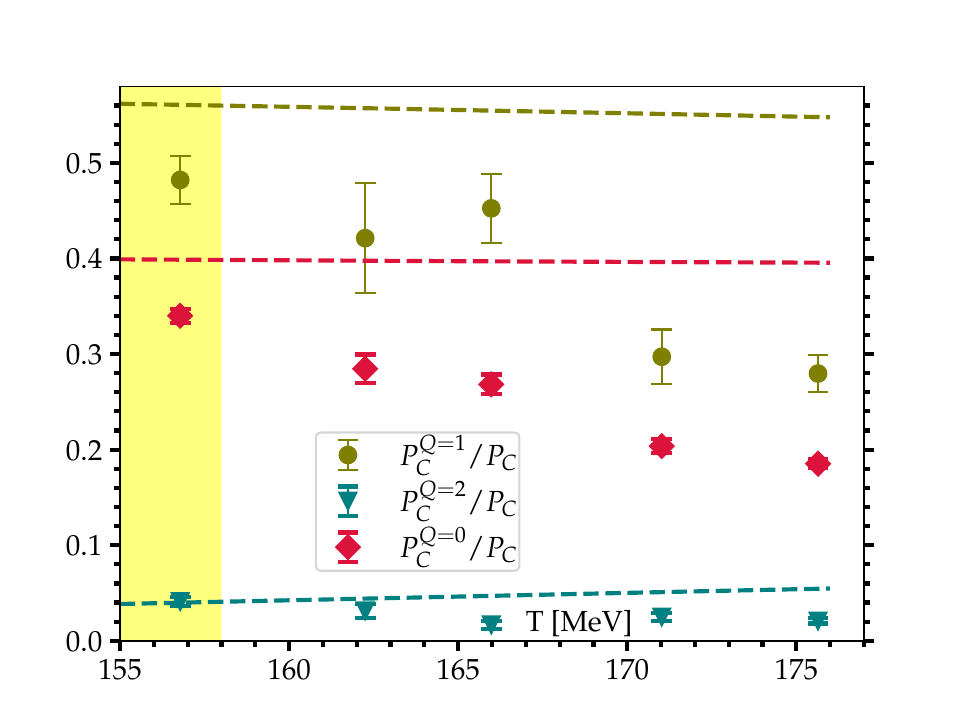}
\caption{Partial charm  pressures for $Q=0,1$ and $2$ sectors normalized by the total
charm pressure as functions of
temperature. Dashed lines show the prediction of the QM-HRG model.
The yellow band represents $T_{pc}$ with its uncertainty.}
\label{fig:PQ=2}
\end{figure}

Additional insight into the charm degrees of freedom above $T_{pc}$ can be obtained 
by considering partial charm pressures corresponding to different electric charge sectors.
Again using Eqs. (\ref{eq:McBc}) and (\ref{eq:Pmodel}) we can determine the partial charm
pressures in different $Q$ and $B$ sectors using the following expressions: 
\begin{align}
	P_{C}^{Q=0}&=\frac{1}{4}\big[4\chi^{C}_{4}-12\chi^{QC}_{13}+11\chi^{QC}_{22}-3\chi^{QC}_{31}\big]\\
	P_{C}^{Q=1}&=-4\chi^{QC}_{13}+8\chi^{QC}_{22}-3\chi^{QC}_{31}\\
	P_{C}^{Q=2}&=\frac{1}{8}\big[2\chi^{QC}_{13}-5\chi^{QC}_{22}+3\chi^{QC}_{31}\big]\\
	P_{C}^{Q=2/3}&=\frac{1}{8}\big[54\chi^{QC}_{13}-81\chi^{QC}_{22}+27\chi^{QC}_{31}\big]\\
	P_{C}^{B=1,Q=2}&=\frac{1}{4}\big[-\chi^{BQC}_{211}+2\chi^{BQC}_{121}-\chi^{BQC}_{112}\big]\\
P_{C}^{B=1,Q=1}&=2\chi^{BQC}_{211}-\chi^{BQC}_{121}
\end{align}
\begin{align}
P_{C}^{B=1,Q=0}&=\frac{1}{2}[2\chi^{BC}_{22}-13\chi^{BQC}_{112}+\chi^{BQC}_{121}+10 \chi^{BQC}_{211}]\\
    P_{C}^{B=1/3,Q=2/3}&=\frac{27}{4}\big[\chi^{BQC}_{112}-\chi^{BQC}_{211}]
\end{align}
We expect the partial pressures for $|Q|=2/3$ and/or ($|Q|=2/3$, $|B|=1/3$) sectors
will agree with $P^C_q$. Using the lattice QCD results for the generalized susceptibilities we
estimated $P_{C}^{Q=2/3}$ and $P_{C}^{B=1/3,Q=2/3}$. The results are shown in Fig. \ref{fig:Pqc}. 
As expected these partial pressures vanish below $T_{pc}$ and
agree with the charm quark pressure $P_q^C$. 
This again shows that 
charm quark degrees of freedom appear
just above the chiral crossover temperature. 
The partial charm pressures with integer values of $Q$ correspond to combinations
of charmed baryon and charmed meson pressures. We find that all these partial pressures differ
significantly from zero above the chiral crossover temperature.
In particular, the partial charm pressure with $|Q|=2$
comes from charmed baryon contributions. In Fig. \ref{fig:PQ=2} we show the corresponding
lattice QCD results. As can be seen,  the partial charm pressures $P_{C}^{|Q|=0,1,2}$ are non-zero above
$T_{pc}$ implying again that charmed hadron-like excitations still exist at
$T\simeq 175$ MeV. At the same time these partial pressures show significant deviations from QM-HRG.

\section{Summary and conclusions}

In this paper we studied the nature of charm degrees of freedom across the chiral
crossover transition using high statistics lattice QCD calculations on generalized
charm susceptibilities. We showed that below the chiral crossover transition the 
generalized susceptibilities agree with HRG model calculations, although
in order to achieve this agreement 
additional charmed baryon states need to be included that are not yet included in the PDG tables,
but are expected to exist based 
on quark model and lattice QCD calculations. 
We thus confirm and corroborate earlier assertions about the existence of additional
charmed hadrons \cite{BAZAVOV2014210}. The high statistics results of generalized susceptibilities also allow to clarify the nature of the charm degrees of freedom above
$T_{pc}$. We find that a charm quark
contribution to the pressure
appears as new degree of freedom just above $T_{pc}$. However, charm quarks 
become the dominant degree of freedom
only at $T> 175$~MeV.
Charmed meson- and baryon-like excitations exist above $T_{pc}$ and
make up half of the contribution to the charmed pressure at temperature $T\simeq 175$ MeV.

The lattice QCD calculations presented in this paper have been performed
at a single value of lattice spacing in temperature units that correspond to
$N_{\tau}=8$ lattices. We find that cutoff effects in the generalized susceptibilities
are insignificant in ratios of different generalized susceptibilities because one of the main sources of these cut-off effects in the generalized susceptibilities is the bare charm quark mass, and while taking a ratio this effects largely cancels.
These cutoff effects can be estimated by considering different prescriptions for
fixing the lines of constant physics for the charm quark mass. While the choice of LCP
has a large influence on the absolute values of generalized susceptibilities the effect cancels
in ratios of generalized susceptibilities. Thus our conclusions will not be affected by discretization errors. All data from our calculations, presented in the figures
of this paper, can be found in Ref. \cite{data}.

\section*{Acknowledgements}
This material is based upon work supported by The U.S. Department of Energy, Office of Science, Office of Nuclear Physics through Contract No.~DE-SC0012704, and within the frameworks of Scientific Discovery through Advanced Computing (SciDAC) award \textit{Fundamental Nuclear Physics at the Exascale and Beyond} and the Topical Collaboration in Nuclear Theory \textit{Heavy-Flavor Theory (HEFTY) for QCD Matter} as well as by the U.S. National Science Foundation under award PHY-2309946.
This work was supported by The Deutsche Forschungsgemeinschaft (DFG, German Research Foundation) - Project number 315477589-TRR 211,
”Strong interaction matter under extreme conditions” and
project number 460248186 (PUNCH4NFDI).

The authors gratefully acknowledge the
computing time and support provided to them on the high-performance computer Noctua 2 at the NHR Center
PC2 under the project name: hpc-prf-cfpd. These are funded by the Federal Ministry of Education
and Research and the state governments participating on the basis of the resolutions of the GWK
for the national high-performance computing at universities (www.nhr-verein.de/unsere-partner).
Numerical calculations have also been performed on the
GPU-cluster at Bielefeld University, Germany. We thank the Bielefeld HPC.NRW team for their support.

All computations in this work were performed using \texttt{SIMULATeQCD} code ~\cite{HotQCD:2023ghu}. All the HRG calculations were performed using the AnalysisToolbox code
developed by the HotQCD Collaboration \cite{Altenkort:2023xxi}.

\bibliographystyle{elsarticle-num} 
\bibliography{refs}

\end{document}